\documentclass[english,a4paper]{article}
\usepackage{dcolumn}
\usepackage{bm}
\usepackage{graphicx}
\usepackage{amsmath}
\usepackage{amsfonts}
\usepackage{amssymb}
\usepackage{amsthm}
\usepackage{amstext}
\usepackage{amsbsy}
\usepackage{amsopn}
\usepackage{amscd}
\usepackage{amsxtra}

\def\openone{\leavevmode\hbox{\small1\kern-3.3pt\normalsize1}}

\usepackage[usenames,dvipsnames]{color}

\begin{document}
\title{Linking the rotation of a rigid body to the Schr\"odinger equation: The quantum tennis racket effect and beyond}
\author{L. Van Damme, D. Leiner, P. Marde\v si\'c\footnote{Institut de Math\'ematiques de Bourgogne, UMR 5584 CNRS-Universit\'e de Bourgogne Franche-Comt\'e, 9 Av. A. Savary, BP 47870 21078 Dijon Cedex, France}, S. J. Glaser\footnote{Department of Chemistry, Technische Universit\"at
M\"unchen, Lichtenbergstrasse 4, D-85747 Garching, Germany}, D. Sugny\footnote{Laboratoire Interdisciplinaire Carnot de
Bourgogne (ICB), UMR 6303 CNRS-Universit\'e Bourgogne-Franche Comt\'e, 9 Av. A.
Savary, BP 47 870, F-21078 Dijon Cedex, France and Institute for Advanced Study, Technische Universit\"at M\"unchen, Lichtenbergstrasse 2 a, D-85748 Garching, Germany, dominique.sugny@u-bourgogne.fr}}

\maketitle

\begin{abstract}
The design of efficient and robust pulse sequences is a fundamental requirement in quantum control. Numerical methods can be used for this purpose, but with relatively little insight into the control mechanism. Here, we show that the free rotation of a classical rigid body plays a fundamental role in the control of two-level quantum systems by means of external electromagnetic pulses. For a state to state transfer, we derive a family of control fields depending upon two free parameters, which allow us to adjust the efficiency, the time and the robustness of the control process. As an illustrative example, we consider the quantum analog of the tennis racket effect, which is a geometric property of any classical rigid body. This effect is demonstrated experimentally for the control of a spin 1/2 particle by using techniques of Nuclear Magnetic Resonance. We also show that the dynamics of a rigid body can be used to implement one-qubit quantum gates. In particular, non-adiabatic geometric quantum phase gates can be realized based on the Montgomery phase of a rigid body. The robustness issue of the gates is discussed.
\end{abstract}

\begin{figure}[!htpb]
\centering
      \includegraphics[width=\linewidth]{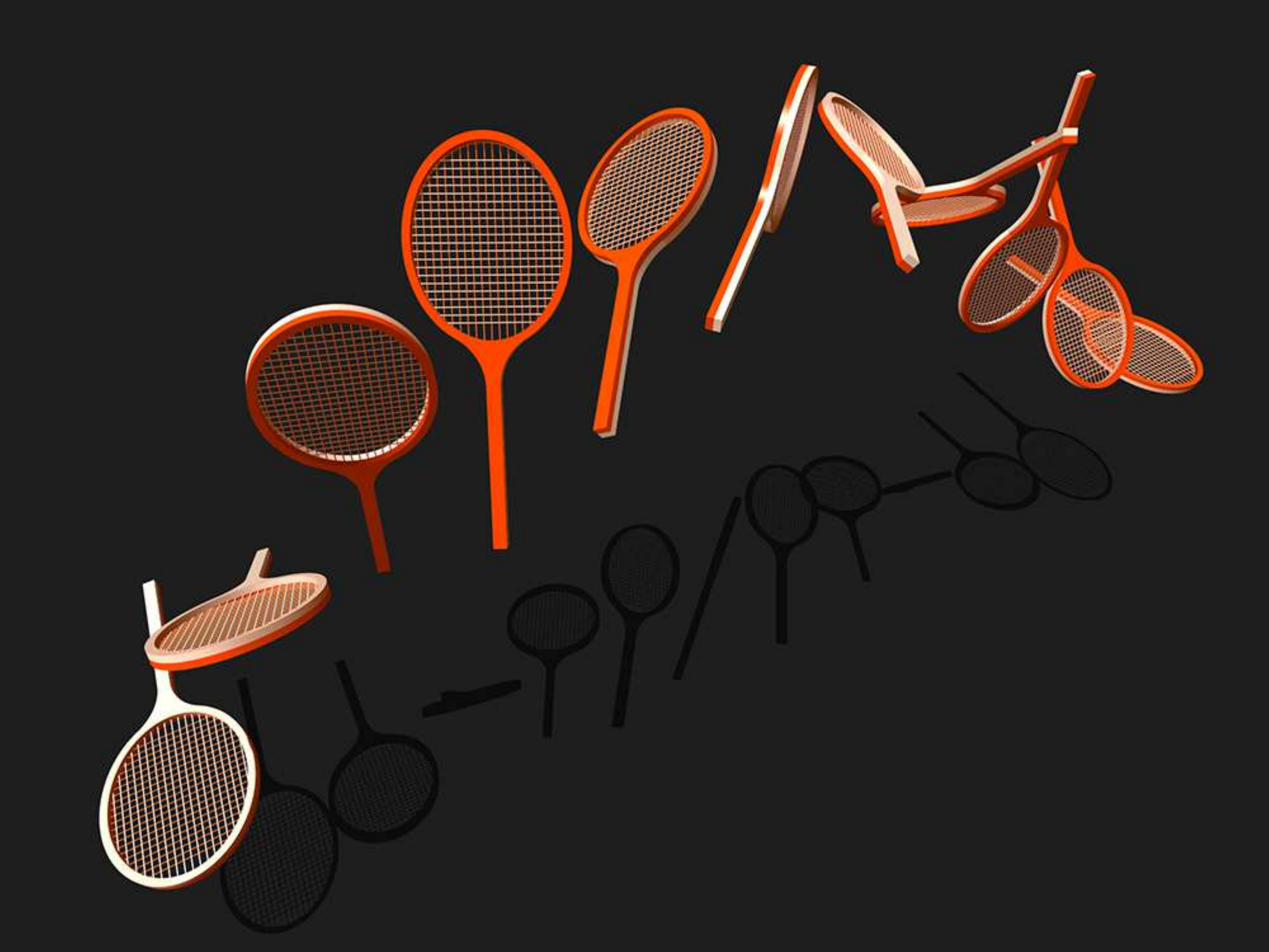}
\caption{(Color online) \textbf{The tennis racket effect}. Illustration of the motion of the tennis racket. Note the flip of the head when the handle makes a $2\pi$ rotation.}
\label{figracket}
\end{figure}

Quantum control is aimed at manipulating dynamical processes at microscopic scales by means of external electromagnetic fields \cite{cat,Brif2010,fidelitycontrol,naturerobust,QSL,prlmintert}. Its successful experimental implementation requires robustness against parameter fluctuations and uncertainties, but also high efficiency in a sufficiently short time to avoid parasitic phenomena such as relaxation. These objectives can be viewed as a crucial prerequisite for a wide range of applications of such techniques in the emerging domain of quantum technologies \cite{cat}. In this setting, numerical algorithms based on optimal control theory \cite{Pontryagin1964} have been developed to realize a given task, while minimizing the control time and accounting for experimental constraints and imperfections \cite{Khaneja2005}. In spite of its efficiency, this approach does not give a clear insight into the control mechanism, which makes it system-dependent and prevents its generalization. The physical understanding of a control process can be extracted from a geometric analysis of the dynamics \cite{Khaneja2001,Lapert2010,shortcut,hoult}. The geometric properties of the corresponding physical effect will ensure its robustness against experimental errors and thus its usefulness \cite{shortcutrobust,shortcutrobust2}. The richness of this geometric approach is illustrated by the Berry phase in quantum mechanics \cite{berryphase,nakahara}. The discovery of the Berry phase led to an impressive amount of studies both in quantum physics and chemistry. Geometric control protocols, resilient to certain types of experimental uncertainties, were developed in quantum computing from this effect \cite{berryqc1,berryqc2}. In this work, we propose to use the study of the free rotation of a rigid body to develop new geometric quantum control strategies. A geometric property, known as the Tennis Racket Effect (TRE) \cite{ashbaugh,vandamme2015}, will be used as an illustrative example to describe this method. This phenomenon occurs in the free rotation of any three-dimensional rigid body \cite{arnold,goldstein,cushman}. It can be easily observed with a tennis racket through the following experimental protocol. We first mark the different faces of the head of the racket. We then take the racket by the handle and throw it in the air so that the handle makes a $2\pi$ rotation. After catching the handle, we observe that the head of the tennis racket has made a flip of $\pi$. This effect can be reproduced for many different rigid bodies and a large range of initial conditions, corresponding to the initial inclination and velocity of the head of the racket, showing thus its robustness, inherent to its geometric character. An illustration of TRE is given in Fig.~\ref{figracket}. A complete mathematical description of TRE was given in a series of papers \cite{cushman,ashbaugh,vandamme2015}. These analyses are based on the fact that the free rotation of a rigid body \cite{cushman} is an integrable system whose trajectories can be derived analytically by using Jacobi elliptic functions \cite{goldstein}. Here, we show that the TRE, and more generally the dynamics of a rigid body, find remarkable applications in the control of two-level quantum systems \cite{Brif2010,cat}. We first obtain a family of control fields based on the TRE allowing to manipulate the state of the system in a robust manner with respect to some experimental uncertainties. Such fields depend on two parameters that can be adjusted to change the time, the efficiency and the robustness of the control process. We introduce the concept of a quantum TRE, which is the analog of the classical motion at the quantum level and we point out its specific quantum properties. The TRE control strategy is demonstrated experimentally on a spin 1/2 particle by using techniques of Nuclear Magnetic Resonance \cite{Levitt08}. We also show that the dynamics of a rigid body allows us to design control fields to realize one-qubit quantum gates. In particular, we focus on the Montgomery phase \cite{montgomery}, a geometric feature of the free rotation of a rigid body, which leads to quantum geometric phase gates \cite{berryqc1}  in the non-adiabatic regime.

A formal equivalence can be established between the free rotation of a rigid body and the dynamics of a spin 1/2 particle, which are governed respectively by the Euler and the Bloch equations. The two systems of differential equations have a similar mathematical structure of the form $\dot{\vec{X}}=H(t)\vec{X}$ where the matrix $H(t)$ can be written as follows:
\begin{equation}\label{eqham}
H(t)=\left(\begin{array}{ccc}
0 & -\Omega_3 & \Omega_2 \\
\Omega_3 & 0 & -\Omega_1 \\
-\Omega_2 & \Omega_1 & 0
\end{array}\right).
\end{equation}
The state of the system is described by the vector $\vec{X}(t)$ and the $\Omega_i$s denote the angular velocities along the $i$-direction, $i=1,2,3$. We refer the reader to Supp. Sec.~I or to textbook references for technical details \cite{arnold,goldstein,Levitt08}. The vector $\vec{X}$ can be identified with the angular momentum $\vec{L}$ of the rigid body (in the frame attached to the racket) or with the Bloch  vector $\vec{M}$ of the spin (in a given rotating frame \cite{Levitt08}). In the classical system, the components of $\vec{L}=(L_1,L_2,L_3)$ can be expressed in terms of the $\Omega_i$s through the principal moments of inertia $I_i$, $L_i=I_i\Omega_i$ (in the principle axis system of the inertia tensor), while in the quantum case, the angular velocities refer to external control fields applied along a given direction. The classical system admits two constants of motion making it Liouville integrable, namely the energy $E=\frac{L_1^2}{2I_1}+\frac{L_2^2}{2I_2}+\frac{L_3^2}{2I_3}$ and the norm of the angular momentum, $L_1^2+L_2^2+L_3^2$, which can be set to 1.  If the control fields applied to the spin are exactly equal to the angular velocities of the rigid body then a one-to-one mapping can be defined between the trajectories of the classical and quantum objects. The moments of inertia can be viewed as additional degrees of freedom used to design control fields with specific properties (see Supp. Sec. II and III). In some experimental applications, only two external fields are available. In this limiting case, an ideal rigid body for which one of the moments of inertia goes to infinity can be considered. Note that the different geometric features of a rigid body are not modified in this limit. In the rest of the paper, we will assume that the three moments of inertia are equal to 1, $1/k^2$ with $k\in ]0,1[$, and $+\infty$.

\noindent \textbf{Classical and quantum tennis racket effects.} Returning back to the dynamical behavior of the tennis racket, two fundamental motions can be considered (See Supp. Sec.~I and IV).

\begin{figure}[!htpb]
\centering
      \includegraphics[width=\linewidth]{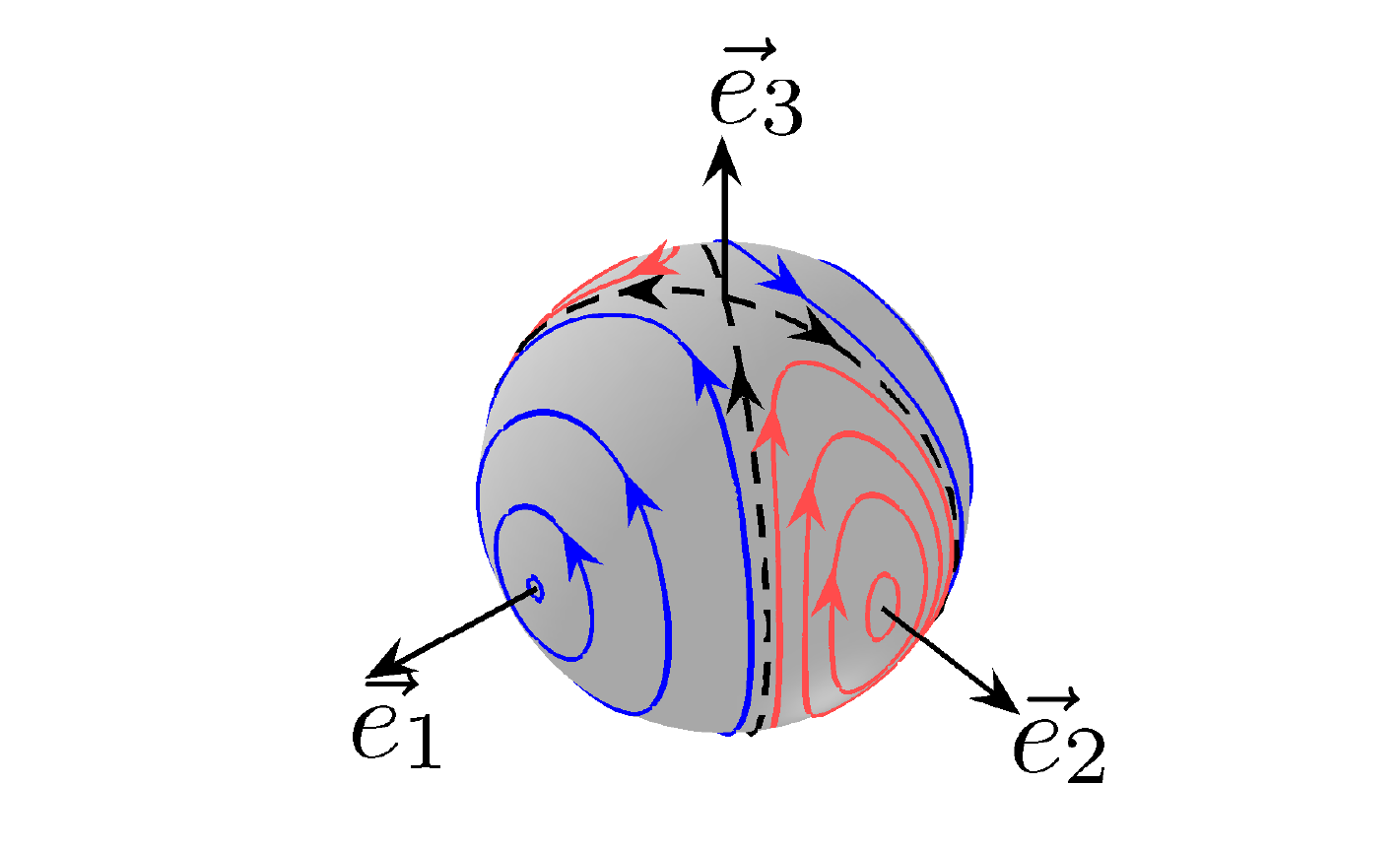}
\caption{(Color online) \textbf{Dynamics of the angular momentum of a three-dimensional rigid body.} Trajectories of the angular momentum $\vec{L}$ of a three-dimensional rigid body in the body-fixed frame $(\vec{e}_1,\vec{e}_2,\vec{e}_3)$. The blue (dark gray) and red (light gray) lines depict respectively the rotating and oscillating trajectories of the angular momentum. The dashed line represents the separatrix. The parameter $k$ is set to 0.5. In the case of a corresponding two-level quantum system, the trajectories represent the dynamics of the Bloch vector in the rotating frame.}
\label{phasespace}
\end{figure}

The first motion is associated with the angular momentum $\vec{L}$ in the reference frame of the racket. During the rotation of the racket, the angular momentum is brought from its initial position to the diametrically opposite one. The different trajectories that can be followed by $\vec{L}$ are displayed in Fig.~\ref{phasespace}. The classical phase space has a simple structure made of a separatrix which is the boundary between two families of trajectories: the rotating and the oscillating ones, each distributed around a stable fixed point \cite{goldstein}. In the example of Fig.~\ref{phasespace} where $\Omega_2(t)=0$, $I_1=1$ and $I_3=1/k^2$, a transfer from the north pole ($\vec{e}_3$) to the south pole ($-\vec{e}_3$) can be achieved on the Bloch sphere by following the separatrix (see Supp. Sec. II). In the quantum mechanical setting, the control requires an infinite time to be performed and corresponds to an Allen-Eberly type pulse sequence of the form \cite{eberly}:
\begin{equation}
\Omega_1(t)=\frac{\pm 1}{\sqrt{1-k^2}}\textrm{sech}(t+t_0);~\Omega_3(t)=\frac{\pm k}{\sqrt{1-k^2}}\textrm{tanh}(t+t_0),
\end{equation}
where $t_0$ is an arbitrary constant time.

The formal equivalence used in this work leads therefore to an insightful geometric interpretation of the Allen-Eberly solution as a singular trajectory of a classical rigid body (see Supp. Sec.~II). In addition to this control strategy, two families of control fields can be derived from TRE. Such solutions, called TRE pulses, correspond to the oscillating or rotating trajectories close to the separatrix, which bring approximately the system from the north to the south pole of the Bloch sphere. Each element of the two sets can be characterized by the parameter $k$ and a small positive constant $\epsilon$, which describes the distance of the trajectory to the separatrix. The two parameters can be chosen to adjust the efficiency, the robustness and the time of the control process. More generally, we show in Supp. Sec.~II by considering the whole range of variations of $\epsilon$ and $k$ that a smooth transition can be established between $\pi$ pulses of constant phase and Allen-Eberly solutions. In the case of Fig.~\ref{phasespace}, the rotation axis of the $\pi$ pulse is associated with one of the two stable fixed points and the Allen-Eberly control with the separatrix. All the other trajectories, and in particular the TRE pulses, represent a compromise between the two solutions.

\begin{figure}[!htpb]
\centering
      \includegraphics[width=\linewidth]{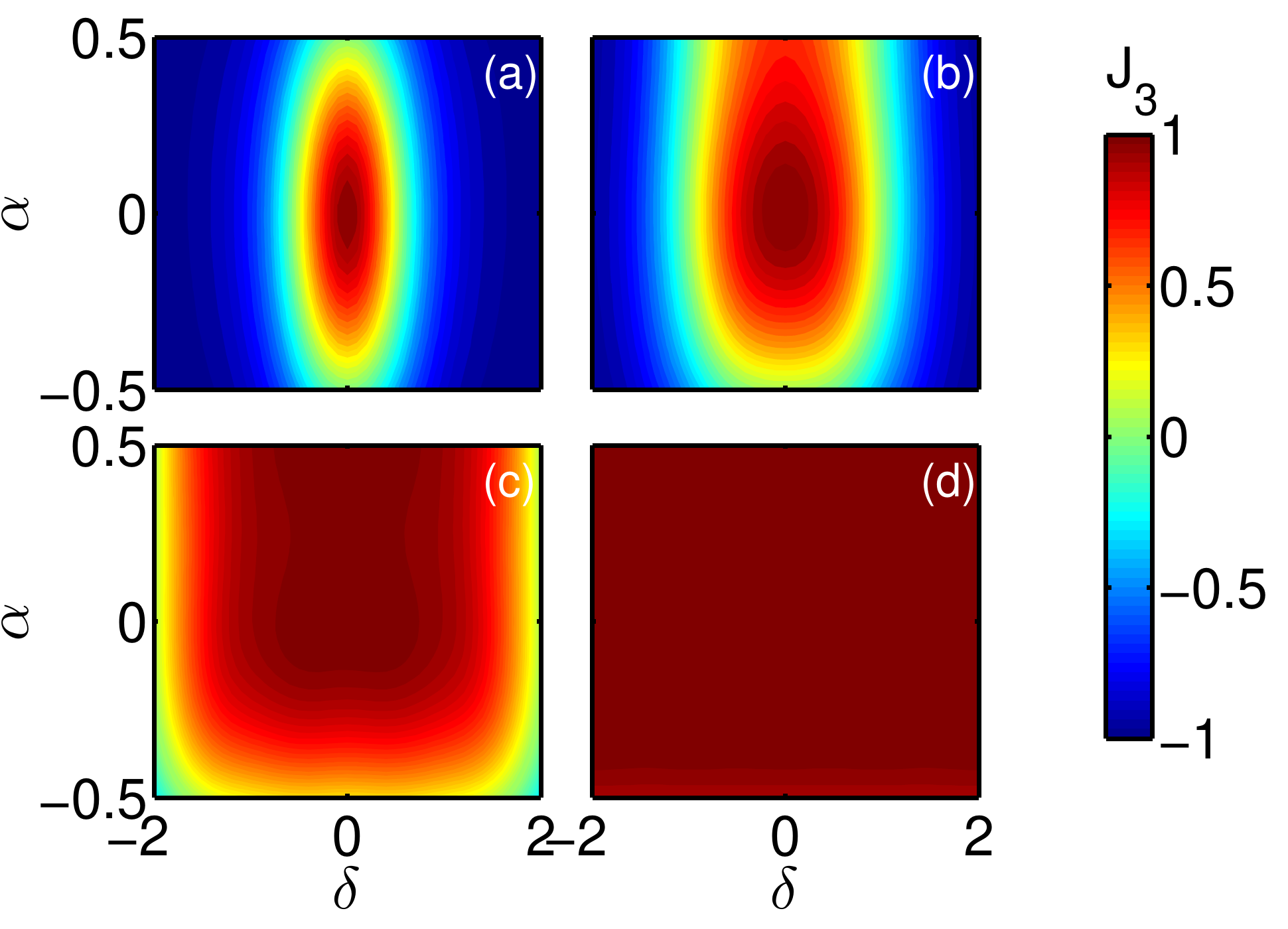}
\caption{(Color online) \textbf{Efficiency and robustness of the Tennis Racket Effect pulses}. The figure of merit $J_3$ is plotted as a function of the $\alpha$ and $\delta$ parameters. In the panels (a), (b), (c) and (d), $k$ is respectively fixed to 0.2, 0.6, 0.9 and 0.99. The parameter $\epsilon$ is set to 0.01 (see Supplementary Sec.~II for details).}
\label{figrobust}
\end{figure}

To evaluate the robust character of the TRE pulse, we consider the control of an ensemble of spins with different offset frequencies $\delta$ and scaling factors $\alpha$ of the amplitude of the control field, the two parameters belonging to intervals fixed by the experimental setup. This description reproduces the standard experimental uncertainties due to the field inhomogeneities that can be encountered in NMR \cite{Levitt08} or in quantum information processing \cite{chuang}. In the numerical simulations, we replace in Eq.~(\ref{eqham}) the three angular velocities by $(1+\alpha) \Omega_1$, $(1+\alpha) \Omega_2$ and $\Omega_3+\delta$. We denote by $t_f$ and $J_3=-M_3(t_f)$, the control time and the figure of merit of the process, respectively. The initial state is the north pole of the Bloch sphere. Figure~\ref{figrobust} shows the efficiency of the TRE pulse. We observe that the robustness of the process changes with the parameter $k$. It can be verified that this property does not depend on $\epsilon$, for $\epsilon$ sufficiently small. The parameter $\epsilon$ affects predominantly the fidelity and the control time of the process. The analytical computations reveal that this time has a logarithmic divergence when $\epsilon$ goes to 0 (see Supp. Sec.~II).

\begin{figure}[!htpb]
\centering
      \includegraphics[width=\linewidth]{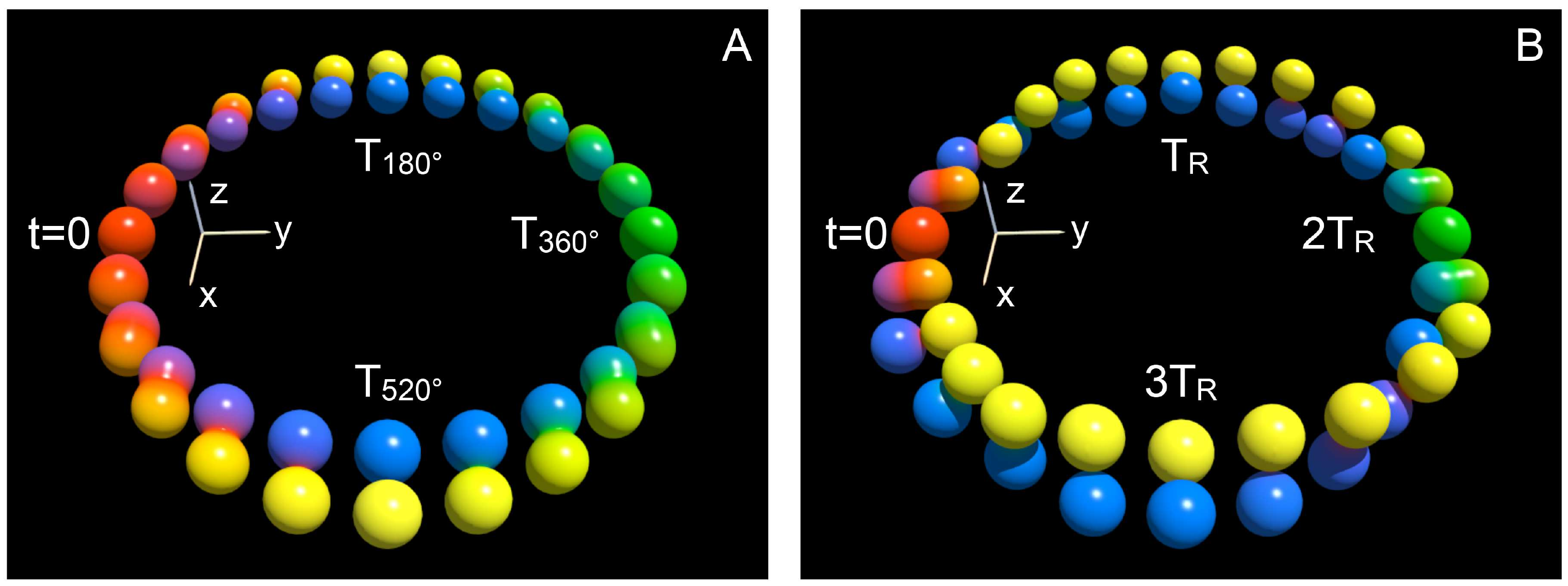}
\caption{(Color online) \textbf{Drops representation of the propagators}. Trajectories of the propagators $U(t)$ for standard rectangular pulses of constant amplitude and phase (A) and by TRE pulses (B) as a function of time. In panels (A) and (B), $T_{180}$ and $T_R$ are pulse durations corresponding to a $180^\circ$- rotation around the $x$- axis. In the DROPS representation \cite{garon}, operators are depicted by complex spherical functions $f(\theta,\phi)=|f(\theta,\phi)|e^{i\eta}$, where for given azimuthal and polar angles $\theta$ and $\phi$, the absolute value $|f(\theta,\phi)|$ is represented by the distance from the origin and the phase angle is color coded ($\eta=0$: red, $\eta=\frac{\pi}{2}$: yellow, $\eta=\pi$: green, $\eta=\frac{3\pi}{2}$: blue). At $t=0$, the propagator is the identity operator \textbf{1} (represented by red spheres), while at (A) $t=T_{360^\circ}=2T_{180^\circ}$ and (B) $t=2T_R$ the propagator is -\textbf{1} (represented by green spheres). In panels (A) and (B), the identity operator is created again at $t=4T_{180^\circ}=T_{720^\circ}$ and $t=4T_R$, respectively.}
\label{figquantumracket}
\end{figure}

A second relevant dynamical process is associated with the motion of the frame attached to the racket with respect to the laboratory frame $(x,y,z)$. Denoting by $R(t)\in SO(3)$, the corresponding rotation matrix at a time $t$, whose dynamics is ruled by the equation $\dot{R}(t)=H(t)R(t)$, the final position of the racket is characterized in the ideal case by (see Supp. Sec.~IV):
\begin{equation}
\label{eq:unitary}
R_f=\left(\begin{array}{ccc} 1 & 0 & 0 \\ 0 & -1 & 0 \\ 0 & 0 & -1\end{array}\right).
\end{equation}
This analogy can be interpreted as a first step towards the implementation of quantum gates, here a NOT gate. However, this transformation is less robust than the one of the angular momentum because the total time of the process has to be perfectly adjusted in order to realize the $2\pi$ rotation of the handle (see below and Supp. Sec.~V). Note that the racket exactly goes back to its initial position after 2 TREs. This geometric phenomenon can be extended to a purely quantum property by using the standard mapping between $SO(3)$ and $SU(2)$. A quantum TRE is then defined from the solution of the Schr\"odinger equation $i\frac{d}{dt}U(t)=\mathcal{H}(t)U(t)$ where $U(t)\in SU(2)$ and $\mathcal{H}$ is the $2\times 2$- Hamiltonian matrix with complex entries corresponding to the Hamiltonian $H(t)$ of Eq.~(\ref{eqham}), which is defined by:
\begin{equation}
\mathcal{H}=\frac{1}{2}\left(\begin{array}{cc} -\Omega_3 & \Omega_1-i\Omega_2 \\ \Omega_1+i\Omega_2 & \Omega_3\end{array}\right).
\end{equation}
We observe that after one TRE, the rotation matrix is given by
\begin{equation}
U_f=(-i)\left(\begin{array}{ccc} 0 & 1 \\ 1 & 0\end{array}\right),
\end{equation}
so that four TREs are needed for the quantum racket to come back to its initial state. A description of this quantum motion is displayed in Fig.~\ref{figquantumracket} by using the DROPS representation of the propagator $U(t)$ \cite{garon}, illustrating the orientation of the effective rotation axis. While in the conventional experiment, this orientation is constant (along the $x$- axis), it follows in the TRE case a twisted trajectory from the $y$- via $z$- to the $x$- axes between $t=0$ and $t=T_R$.

\begin{figure}[!htpb]
  \centering
  \includegraphics[width=\linewidth]{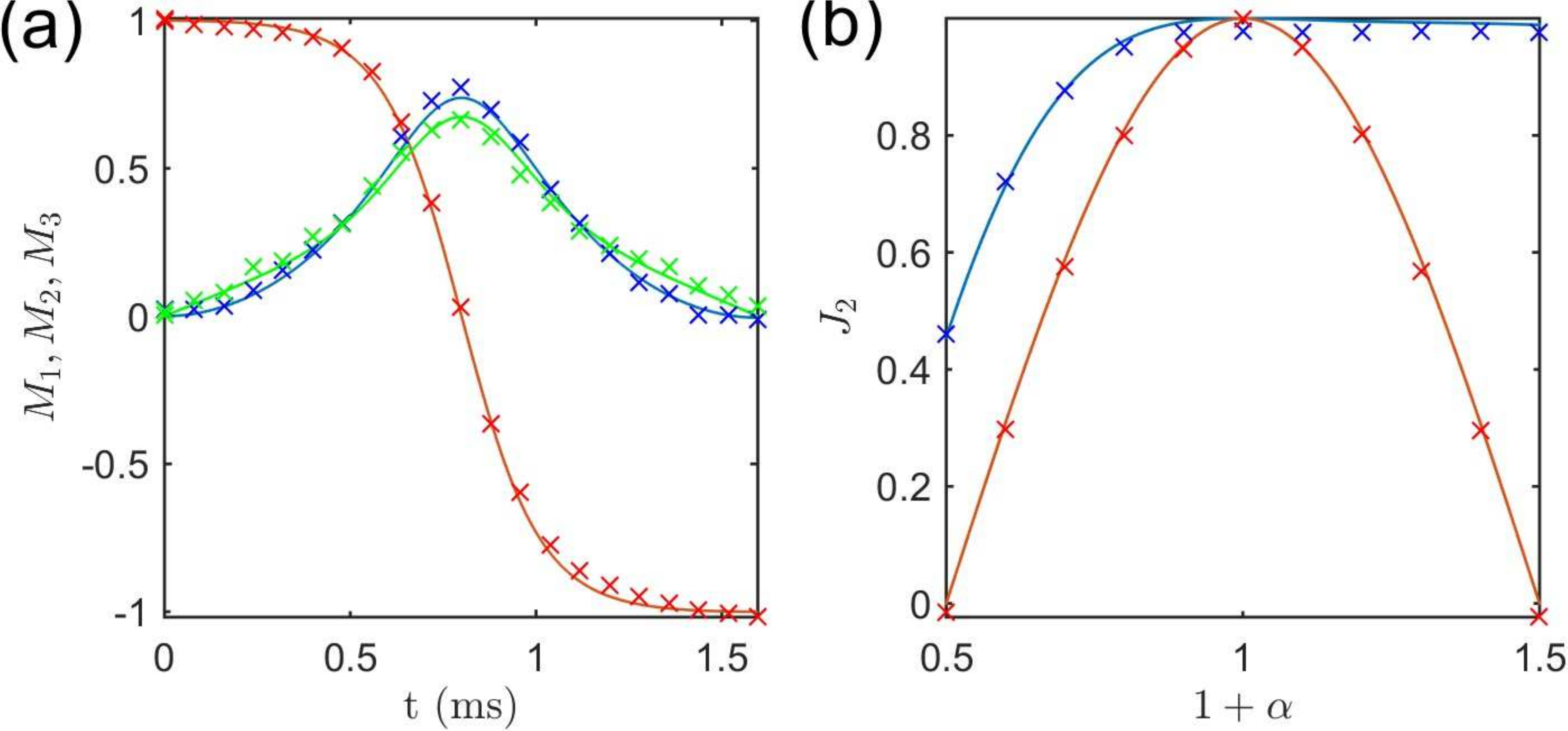}
  \caption{\label{fig:experiment} (Color online) \textbf{Experimental implementation of the TRE pulse.} Experimental details are given in the Supplementary Sec. VI. The panels (a) and (b) represent respectively the trajectories of the components of the Bloch vector ($2$: red or dark gray, $3$: blue or black, $1$: green or light gray) and the robustness of $J_2$ with respect to the $\alpha$- parameter. A rectangular $\pi$ pulse is used in panel (b) for comparison (red or dark gray).}
\end{figure}

\noindent\textbf{Experimental implementation.} We now show by using NMR techniques the experimental performance of a TRE pulse to realize a state to state transfer from $(0,1,0)$ to $(0,-1,0)$. The evolution of the Bloch vector and the robustness of the shaped pulse with respect to the scaling factor $\alpha$ are displayed in Fig.~\ref{fig:experiment}. For the robustness experiments, we scaled the amplitude with factors ranging from $\alpha=-0.5$ to $\alpha=0.5$ in $11$ steps. The total pulse duration was set to $0.448$ ms. The figure of merit $J_2$ is here defined by $J_2=-M_2(t_f)$. A reasonable match is found between theoretical and experimental results (See Supp. Sec.~VI for a discussion about the experimental errors).

\begin{figure}[h!]
\centering
\includegraphics[width=\linewidth]{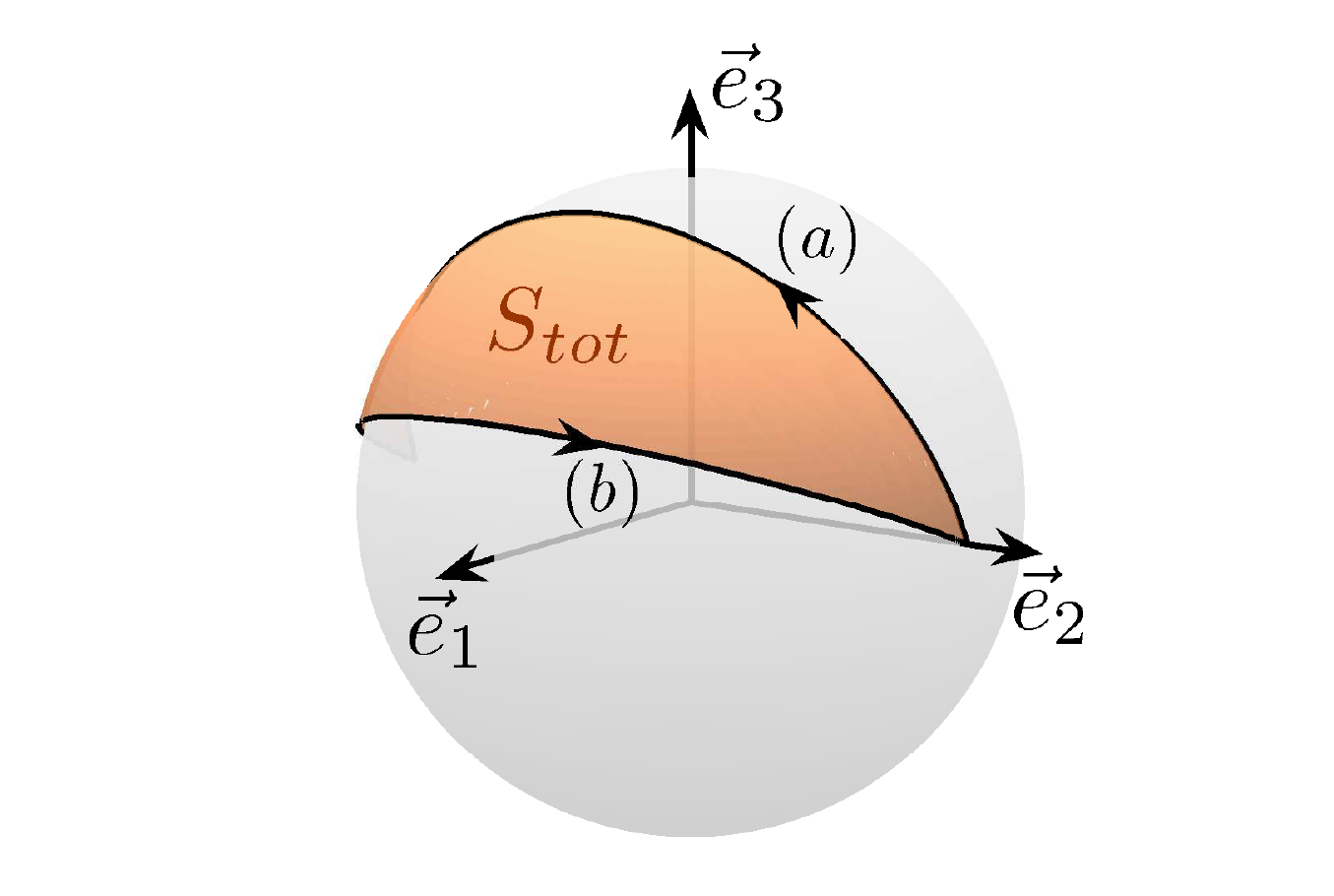}
\caption{(Color online) \textbf{Quantum gate based on the Montgomery phase}. Cyclic process on the Bloch sphere for implementing a $\pi/2$ phase gate. The loop is the concatenation of the trajectories (a) and (b) which are defined by different values of $k$ (see Supplementary Sec. V). The brown (gray) surface is the total geometric phase at the end of the process i.e. $S_{tot}=\pi/2$.\label{fig6}}
\end{figure}

\noindent \textbf{Implementation of one-qubit gates.} The correspondence between the free rotation of a rigid body and the dynamics of a spin 1/2 particle also provides novel control strategies in quantum computing~\cite{chuang}. In Supp. Sec.~V, we show how to implement the Hadamard gate, and more generally any one-qubit quantum gate. Here, we focus on the case of the geometric quantum phase gates~\cite{sjoqvist}, that can be realized by using a geometric feature of the free rotation of a rigid body, namely the Montgomery phase~\cite{montgomery}. This phase can be defined by considering one period of the time evolution of the angular momentum $\vec{L}$ in the body-fixed frame. During this motion, the laboratory frame rotates about $\vec{L}$ by an angle, the Montgomery phase. This phase can be expressed as the sum of a dynamical and a geometric contribution~\cite{berryphase}, this latter being given by $S$, the solid angle swept out by the angular momentum vector (see Supp. Sec. V). One of the main difficulties to realize geometric phase gates is to find a way to cancel the dynamical contribution of the phase in order to obtain a robust control protocol. Different techniques have been proposed up to date \cite{sjoqvist,berryqc1}. Geometric phase gates can be implemented in the adiabatic regime~\cite{berryqc1}, but it is possible to generalize this process to consider non-adiabatic cyclic evolution~\cite{geometricgateNA1,geometricgateNA2}, which is crucial to avoid decoherence effects. Only very simple motions, such as a circle on the Bloch sphere, were proposed. Using the Montgomery phase and the dynamics of a rigid body, this idea can be considerably extended, as illustrated in Fig.~\ref{fig6}. In the example of Fig.~\ref{fig6}, the loop on the Bloch sphere is the concatenation of two trajectories close to the separatrices with different values of $k$, which are adjusted to cancel the dynamical phases (see Supp. Sec. V). From this control process, any geometric phase gate can be implemented in the non-adiabatic regime. More generally, the dynamical phase is also at the origin of the relatively low robustness of the one-qubit quantum gates realized based on the dynamics of a rigid body. This property can be greatly improved by considering a generalization of the BIR- pulses used in NMR~\cite{BR1,BR2}. This control strategy consists in the concatenation of two (or more) pulses chosen so that the global dynamical phase is cancelled. The example of a NOT gate is described in Supp. Sec.~V.

\noindent \textbf{Discussion.} By using the formal analogy between the free rotation of a rigid body and the dynamics of a spin 1/2 particle, we have derived a new family of control fields able to realize either a state to state transfer or a specific unitary transformation in a two-level quantum system. As demonstrated in this paper, the derived pulses have an explicit and relatively simple form, which is therefore easily implementable experimentally. Note that a Matlab code computing the trajectories of a rigid body and of the corresponding Bloch vector is provided in Supp. Sec.~VII.

The results of this work pave the way to other analyses using the same kind of equivalence. The applicability of this analogy beyond simple two-level quantum systems, such as in a chain of three coupled spins \cite{khaneja,VZGS14}, shows the general interest of this approach. Following the method proposed in Ref. \cite{khaneja}, the control fields derived from the dynamics of a rigid body could also be used as a building block to realize a CNOT gate in this system. Another possible direction is the generalization of this study to $SO(n)$, with $n>3$, for instance in the integrable case of the Manakov top \cite{perelomov}.

\vspace{1cm}

\noindent\textbf{ACKNOWLEDGMENT}\\
S.J. Glaser acknowledges support from the DFG (Gl 203/7-2). D. Sugny and S. J. Glaser acknowledge support from the ANR-DFG research program Explosys (ANR-14-CE35-0013-01; Gl203/9-1). This work has been done with the support of the Technische Universit\"at M\"unchen – Institute for Advanced Study, funded by the German Excellence Initiative and the European Union Seventh Framework Programme under grant agreement 291763. Experiments were performed at the Bavarian NMR center at TU M\"unchen. P. Marde\v si\'c acknowledges support from LAISLA (project funded by FONCICYT) and the PREI project (funded by UNAM DGAPA).\\
\noindent \textbf{Author contributions}\\
All authors contributed to the design and interpretation of the
presented work. Numerical computations have been done by L. V. D., the construction of the experiment and the acquisition of the data were performed by D. L. and S. J. G.\\
\noindent \textbf{Additional information}\\
Supplementary information accompanies this paper.\\
Competing financial interests: The authors declare no competing
financial interests.\\
\noindent \textbf{Author information}\\
Correspondence and requests for materials should be addressed to
D. S. (email: dominique.sugny@u-bourgogne.fr) and S. J. G. (email: glaser@tum.de).\\ \\

\vspace{1cm}
\newpage

\begin{thebibliography}{32}
\expandafter\ifx\csname natexlab\endcsname\relax\def\natexlab#1{#1}\fi
\expandafter\ifx\csname bibnamefont\endcsname\relax
  \def\bibnamefont#1{#1}\fi
\expandafter\ifx\csname bibfnamefont\endcsname\relax
  \def\bibfnamefont#1{#1}\fi
\expandafter\ifx\csname citenamefont\endcsname\relax
  \def\citenamefont#1{#1}\fi
\expandafter\ifx\csname url\endcsname\relax
  \def\url#1{\texttt{#1}}\fi
\expandafter\ifx\csname urlprefix\endcsname\relax\def\urlprefix{URL }\fi
\providecommand{\bibinfo}[2]{#2}
\providecommand{\eprint}[2][]{\url{#2}}

\bibitem{cat}
\bibinfo{author}{\bibfnamefont{S.~J.} \bibnamefont{Glaser}},
  \bibinfo{author}{\bibfnamefont{U.}~\bibnamefont{Boscain}},
  \bibinfo{author}{\bibfnamefont{T.}~\bibnamefont{Calarco}},
  \bibinfo{author}{\bibfnamefont{C.~P.} \bibnamefont{Koch}},
  \bibinfo{author}{\bibfnamefont{W.~K.} \bibnamefont{K\"ockenberger}},
  \bibinfo{author}{\bibfnamefont{R.}~\bibnamefont{Kosloff}},
  \bibinfo{author}{\bibfnamefont{I.}~\bibnamefont{Kuprov}},
  \bibinfo{author}{\bibfnamefont{B.}~\bibnamefont{Luy}},
  \bibinfo{author}{\bibfnamefont{S.}~\bibnamefont{Schirmer}},
  \bibinfo{author}{\bibfnamefont{T.}~\bibnamefont{Schulte-Herbr\"uggen}},
  \bibnamefont{et~al.},
  \emph{\bibinfo{title}{Training Schr\"odinger's cat: Quantum optimal control}},
  \bibinfo{journal}{Eur. Phys. J. D}
  \textbf{\bibinfo{volume}{69}}, \bibinfo{pages}{279} (\bibinfo{year}{2015}).

\bibitem{Brif2010}
\bibinfo{author}{\bibfnamefont{C.}~\bibnamefont{Brif}},
  \bibinfo{author}{\bibfnamefont{R.}~\bibnamefont{Chakrabarti}},
  \bibnamefont{and} \bibinfo{author}{\bibfnamefont{H.}~\bibnamefont{Rabitz}},
   \emph{\bibinfo{title}{Control of quantum phenomena: Past, present and future}},
  \bibinfo{journal}{New J. Phys.} \textbf{\bibinfo{volume}{12}},
  \bibinfo{pages}{075008} (\bibinfo{year}{2010}).

\bibitem{fidelitycontrol}
\bibinfo{author}{\bibfnamefont{M.~G.} \bibnamefont{Bason}},
  \bibinfo{author}{\bibfnamefont{M.}~\bibnamefont{Viteau}},
  \bibinfo{author}{\bibfnamefont{N.}~\bibnamefont{Malossi}},
  \bibinfo{author}{\bibfnamefont{P.}~\bibnamefont{Huillery}},
  \bibinfo{author}{\bibfnamefont{E.}~\bibnamefont{Aeimondo}},
  \bibinfo{author}{\bibfnamefont{D.}~\bibnamefont{Ciampini}},
  \bibinfo{author}{\bibfnamefont{R.}~\bibnamefont{Fazio}},
  \bibinfo{author}{\bibfnamefont{V.}~\bibnamefont{Giovannetti}},
  \bibinfo{author}{\bibfnamefont{R.}~\bibnamefont{Mannella}}, \bibnamefont{and}
  \bibinfo{author}{\bibfnamefont{O.}~\bibnamefont{Morsch}},
  \emph{\bibinfo{title}{High-fidelity quantum driving}},
  \bibinfo{journal}{Nature Phys.} \textbf{\bibinfo{volume}{8}},
  \bibinfo{pages}{147} (\bibinfo{year}{2012}).

\bibitem{naturerobust}
\bibinfo{author}{\bibfnamefont{D.}~\bibnamefont{Leibfried}},
  \bibinfo{author}{\bibfnamefont{B.}~\bibnamefont{DeMarco}},
  \bibinfo{author}{\bibfnamefont{V.}~\bibnamefont{Meyer}},
  \bibinfo{author}{\bibfnamefont{D.}~\bibnamefont{Lucas}},
  \bibinfo{author}{\bibfnamefont{M.}~\bibnamefont{Barrett}},
  \bibinfo{author}{\bibfnamefont{J.}~\bibnamefont{Britton}},
  \bibinfo{author}{\bibfnamefont{W.~M.} \bibnamefont{Itano}},
  \bibinfo{author}{\bibfnamefont{B.}~\bibnamefont{Jelenkovic}},
  \bibinfo{author}{\bibfnamefont{C.}~\bibnamefont{Langer}},
  \bibinfo{author}{\bibfnamefont{T.}~\bibnamefont{Rosenband}},
  \bibnamefont{et~al.},
  \emph{\bibinfo{title}{Experimental demonstration of a robust, high-fidelity geometric two ion-qubit phase gate}},
  \bibinfo{journal}{Nature}
  \textbf{\bibinfo{volume}{422}}, \bibinfo{pages}{412} (\bibinfo{year}{2003}).

\bibitem{QSL}
\bibinfo{author}{\bibfnamefont{J.}~\bibnamefont{Sorensen}},
  \bibinfo{author}{\bibfnamefont{M.}~\bibnamefont{Pedersen}},
  \bibinfo{author}{\bibfnamefont{M.}~\bibnamefont{Munch}},
  \bibinfo{author}{\bibfnamefont{P.}~\bibnamefont{Haikka}},
  \bibinfo{author}{\bibfnamefont{J.}~\bibnamefont{Jensen}},
  \bibinfo{author}{\bibfnamefont{T.}~\bibnamefont{Planke}},
  \bibinfo{author}{\bibfnamefont{M.} \bibnamefont{Andreasen}},
  \bibinfo{author}{\bibfnamefont{M.}~\bibnamefont{Gajdacz}},
  \bibinfo{author}{\bibfnamefont{K.}~\bibnamefont{Molmer}},
  \bibinfo{author}{\bibfnamefont{A.}~\bibnamefont{Lieberoth}},
  \bibnamefont{and}
  \bibinfo{author}{\bibfnamefont{J.}~\bibnamefont{Sherson}},
  \emph{\bibinfo{title}{Exploring the Quantum Speed Limit with Computer Games}},
  \bibinfo{journal}{Nature}
  \textbf{\bibinfo{volume}{532}}, \bibinfo{pages}{210} (\bibinfo{year}{2016}).

\bibitem{prlmintert}
\bibinfo{author}{\bibfnamefont{T.}~\bibnamefont{N\"obauer}},
  \bibinfo{author}{\bibfnamefont{A.}~\bibnamefont{Angerer}},
  \bibinfo{author}{\bibfnamefont{B.}~\bibnamefont{Bartels}},
  \bibinfo{author}{\bibfnamefont{M.}~\bibnamefont{Trupke}},
  \bibinfo{author}{\bibfnamefont{S.}~\bibnamefont{Rotter}},
  \bibinfo{author}{\bibfnamefont{J.}~\bibnamefont{Schmiedmayer}},
  \bibinfo{author}{\bibfnamefont{F.}~\bibnamefont{Mintert}}, \bibnamefont{and}
  \bibinfo{author}{\bibfnamefont{J.}~\bibnamefont{Majer}},
  \emph{\bibinfo{title}{Smooth Optimal Quantum Control for Robust Solid-State Spin Magnetometry}},
  \bibinfo{journal}{Phys. Rev. Lett.} \textbf{\bibinfo{volume}{115}},
  \bibinfo{pages}{190801} (\bibinfo{year}{2015}).

\bibitem{Pontryagin1964}
\bibinfo{author}{\bibfnamefont{L.~S.} \bibnamefont{Pontryagin}},
  \bibinfo{author}{\bibfnamefont{V.~G.} \bibnamefont{Bol'tanskii}},
  \bibinfo{author}{\bibfnamefont{R.~S.} \bibnamefont{Gamkrelidze}},
  \bibnamefont{and} \bibinfo{author}{\bibfnamefont{E.~F.}
  \bibnamefont{Mischenko}}, \emph{\bibinfo{title}{The Mathematical Theory of
  Optimal Processes}} (\bibinfo{publisher}{Pergamon Press, New York},
  \bibinfo{year}{1964}).

\bibitem{Khaneja2005}
\bibinfo{author}{\bibfnamefont{N.}~\bibnamefont{Khaneja}},
  \bibinfo{author}{\bibfnamefont{T.}~\bibnamefont{Reiss}},
  \bibinfo{author}{\bibfnamefont{C.}~\bibnamefont{Kehlet}},
  \bibinfo{author}{\bibfnamefont{T.}~\bibnamefont{Schulte-{H}erbr\"{u}ggen}},
  \bibnamefont{and} \bibinfo{author}{\bibfnamefont{S.~J.}
  \bibnamefont{Glaser}},
  \emph{\bibinfo{title}{Optimal control of coupled spin dynamics: design of {NMR} pulse sequences
	by gradient ascent algorithms}},
    \bibinfo{journal}{J. Magn. Reson.}
  \textbf{\bibinfo{volume}{172}}, \bibinfo{pages}{296} (\bibinfo{year}{2005}).

\bibitem{Khaneja2001}
\bibinfo{author}{\bibfnamefont{N.}~\bibnamefont{Khaneja}},
  \bibinfo{author}{\bibfnamefont{R.}~\bibnamefont{Brockett}}, \bibnamefont{and}
  \bibinfo{author}{\bibfnamefont{S.~J.} \bibnamefont{Glaser}},
  \emph{\bibinfo{title}{Time optimal control in spin systems}},
  \bibinfo{journal}{Phys. Rev. A} \textbf{\bibinfo{volume}{63}},
  \bibinfo{pages}{032308} (\bibinfo{year}{2001}).

\bibitem{Lapert2010}
\bibinfo{author}{\bibfnamefont{M.}~\bibnamefont{Lapert}},
  \bibinfo{author}{\bibfnamefont{Y.}~\bibnamefont{Zhang}},
  \bibinfo{author}{\bibfnamefont{M.}~\bibnamefont{Braun}},
  \bibinfo{author}{\bibfnamefont{S.~J.} \bibnamefont{Glaser}},
  \bibnamefont{and} \bibinfo{author}{\bibfnamefont{D.}~\bibnamefont{Sugny}},
  \emph{\bibinfo{title}{Singular Extremals for the Time-Optimal Control of Dissipative Spin
	1/2 Particles}},
  \bibinfo{journal}{Phys. Rev. Lett.} \textbf{\bibinfo{volume}{104}},
  \bibinfo{pages}{083001} (\bibinfo{year}{2010}).

\bibitem{shortcut}
\bibinfo{author}{\bibfnamefont{X.}~\bibnamefont{Chen}},
  \bibinfo{author}{\bibfnamefont{I.}~\bibnamefont{Lizuain}},
  \bibinfo{author}{\bibfnamefont{A.}~\bibnamefont{Ruschhaupt}},
  \bibinfo{author}{\bibfnamefont{D.}~\bibnamefont{Gu\'ery-Odelin}},
  \bibnamefont{and} \bibinfo{author}{\bibfnamefont{J.~G.} \bibnamefont{Muga}},
  \emph{\bibinfo{title}{Shortcut to Adiabatic Passage in Two- and Three-Level Atoms}},
  \bibinfo{journal}{Phys. Rev. Lett.} \textbf{\bibinfo{volume}{105}},
  \bibinfo{pages}{123003} (\bibinfo{year}{2010}).

\bibitem{hoult}
\bibinfo{author}{\bibfnamefont{M.~S.} \bibnamefont{Silver}},
  \bibinfo{author}{\bibfnamefont{R.~I.} \bibnamefont{Joseph}},
  \bibinfo{author}{\bibfnamefont{C.-N.} \bibnamefont{Chen}},
  \bibinfo{author}{\bibfnamefont{V.~J.} \bibnamefont{Sank}}, \bibnamefont{and}
  \bibinfo{author}{\bibfnamefont{D.~I.} \bibnamefont{Hoult}},
   \emph{\bibinfo{title}{Selective population inversion in NMR}},
  \bibinfo{journal}{Nature} \textbf{\bibinfo{volume}{310}},
  \bibinfo{pages}{681} (\bibinfo{year}{1984}).

\bibitem{shortcutrobust}
\bibinfo{author}{\bibfnamefont{A.} \bibnamefont{Ruschhaupt}},
  \bibinfo{author}{\bibfnamefont{X.} \bibnamefont{Chen}},
  \bibinfo{author}{\bibfnamefont{D.} \bibnamefont{Alonso}},
  \bibnamefont{and}
  \bibinfo{author}{\bibfnamefont{J.~G.} \bibnamefont{Muga}},
   \emph{\bibinfo{title}{Optimally robust shortcuts to population inversion in two-level quantum systems}},
  \bibinfo{journal}{New J. Phys.} \textbf{\bibinfo{volume}{14}},
  \bibinfo{pages}{093040} (\bibinfo{year}{2012}).

\bibitem{shortcutrobust2}
\bibinfo{author}{\bibfnamefont{D.} \bibnamefont{Daems}},
  \bibinfo{author}{\bibfnamefont{A.} \bibnamefont{Ruschhaupt}},
  \bibinfo{author}{\bibfnamefont{D.} \bibnamefont{Sugny}},
  \bibnamefont{and}
  \bibinfo{author}{\bibfnamefont{S.} \bibnamefont{Gu\'erin}},
   \emph{\bibinfo{title}{Robust quantum control by a single-shot shaped pulse}},
  \bibinfo{journal}{Phys. Rev. Lett.} \textbf{\bibinfo{volume}{111}},
  \bibinfo{pages}{050404} (\bibinfo{year}{2013}).


\bibitem{berryphase}
\bibinfo{author}{\bibfnamefont{A.}~\bibnamefont{Bohm}},
  \bibinfo{author}{\bibfnamefont{H.}~\bibnamefont{Mostafazadeh}},
  \bibinfo{author}{\bibfnamefont{Q.}~\bibnamefont{Koizumi}}, \bibnamefont{and}
  \bibinfo{author}{\bibfnamefont{J.}~\bibnamefont{Zwanziger}},
  \emph{\bibinfo{title}{The geometric phase in quantum systems}}
  (\bibinfo{publisher}{Springer, Berlin}, \bibinfo{year}{2003}).

\bibitem{nakahara}
\bibinfo{author}{\bibfnamefont{M.}~\bibnamefont{Nakahara}},
  \emph{\bibinfo{title}{Geometry, topology and physics}}
  (\bibinfo{publisher}{Institute of physics publishing},
  \bibinfo{address}{Bristol and Philadelphia}, \bibinfo{year}{1990}).

\bibitem{berryqc1}
\bibinfo{author}{\bibfnamefont{J.~A.} \bibnamefont{Jones}},
  \bibinfo{author}{\bibfnamefont{V.}~\bibnamefont{Vedral}},
  \bibinfo{author}{\bibfnamefont{A.}~\bibnamefont{Ekert}}, \bibnamefont{and}
  \bibinfo{author}{\bibfnamefont{G.}~\bibnamefont{Castagnoli}},
  \emph{\bibinfo{title}{Geometric quantum computation using Nuclear Magnetic Resonance}},
  \bibinfo{journal}{Nature} \textbf{\bibinfo{volume}{43}}, \bibinfo{pages}{869}
  (\bibinfo{year}{2000}{\natexlab{a}}).

\bibitem{berryqc2}
\bibinfo{author}{\bibfnamefont{A.~A.} \bibnamefont{Abdumalikov}},
  \bibinfo{author}{\bibfnamefont{J.~M.} \bibnamefont{Fink}},
  \bibinfo{author}{\bibfnamefont{K.}~\bibnamefont{Juliusson}},
  \bibinfo{author}{\bibfnamefont{M.}~\bibnamefont{Pechal}},
  \bibinfo{author}{\bibfnamefont{S.}~\bibnamefont{Berger}},
  \bibinfo{author}{\bibfnamefont{A.}~\bibnamefont{Wallraff}}, \bibnamefont{and}
  \bibinfo{author}{\bibfnamefont{S.}~\bibnamefont{Filipp}},
  \emph{\bibinfo{title}{Experimental realization of non-Abelian non-adiabatic geometric gates}},
  \bibinfo{journal}{Nature} \textbf{\bibinfo{volume}{496}}, \bibinfo{pages}{482}
  (\bibinfo{year}{2013}).

\bibitem{ashbaugh}
\bibinfo{author}{\bibfnamefont{M.~S.} \bibnamefont{Ashbaugh}},
  \bibinfo{author}{\bibfnamefont{C.~C.} \bibnamefont{Chiconce}},
  \bibnamefont{and} \bibinfo{author}{\bibfnamefont{R.~H.}
  \bibnamefont{Cushman}},
  \emph{\bibinfo{title}{The twisting tennis racket}},
  \bibinfo{journal}{J. Dyn. Diff. Eq.}
  \textbf{\bibinfo{volume}{3}}, \bibinfo{pages}{67} (\bibinfo{year}{1991}).

\bibitem{vandamme2015}
\bibinfo{author}{\bibfnamefont{L.}~\bibnamefont{Van~Damme}},
  \bibinfo{author}{\bibfnamefont{P.}~\bibnamefont{Mardesic}}, \bibnamefont{and}
  \bibinfo{author}{\bibfnamefont{D.}~\bibnamefont{Sugny}},
  \emph{\bibinfo{title}{The tennis racket effect in a three-dimensional rigid body}},
  \bibinfo{journal}{Physica D}
  \textbf{\bibinfo{volume}{338}}, \bibinfo{pages}{17} (\bibinfo{year}{2017}).

\bibitem{arnold}
\bibinfo{author}{\bibfnamefont{V.~I.} \bibnamefont{Arnold}},
  \emph{\bibinfo{title}{Mathematical methods of classical mechanics}}
  (\bibinfo{publisher}{Springer-Verlag, New York}, \bibinfo{year}{1989}).

\bibitem{goldstein}
\bibinfo{author}{\bibfnamefont{H.}~\bibnamefont{Goldstein}},
  \emph{\bibinfo{title}{Classical mechanics}}
  (\bibinfo{publisher}{Addison-Wesley, Reading, M.A.}, \bibinfo{year}{1950}).

\bibitem{cushman}
\bibinfo{author}{\bibfnamefont{R.~H.} \bibnamefont{Cushman}} \bibnamefont{and}
  \bibinfo{author}{\bibfnamefont{L.~M.} \bibnamefont{Bates}},
  \emph{\bibinfo{title}{Global aspects of classical integrable systems}}
  (\bibinfo{publisher}{Birkh\"auser, Berlin}, \bibinfo{year}{1997}).

\bibitem{Levitt08}
\bibinfo{author}{\bibfnamefont{M.~H.} \bibnamefont{Levitt}},
  \emph{\bibinfo{title}{Spin Dynamics: Basics of Nuclear Magnetic Resonance}}
  (\bibinfo{publisher}{Wiley, New York}, \bibinfo{year}{2008}).

\bibitem{montgomery}
\bibinfo{author}{\bibfnamefont{R.}~\bibnamefont{Montgomery}},
 \emph{\bibinfo{title}{How much does the rigid body rotate ? A Berry's phase from the 18th century}},
  \bibinfo{journal}{Am. J. Phys.} \textbf{\bibinfo{volume}{59}},
  \bibinfo{pages}{394} (\bibinfo{year}{1991}).


\bibitem{eberly}
\bibinfo{author}{\bibfnamefont{L.}~\bibnamefont{Allen}} \bibnamefont{and}
  \bibinfo{author}{\bibfnamefont{J.~H.} \bibnamefont{Eberly}},
  \emph{\bibinfo{title}{Optical resonance and two-level atoms}}
  (\bibinfo{publisher}{Wiley, New York}, \bibinfo{year}{1975}).

\bibitem{chuang}
\bibinfo{author}{\bibfnamefont{M.~A.} \bibnamefont{Nielsen}} \bibnamefont{and}
  \bibinfo{author}{\bibfnamefont{I.~L.} \bibnamefont{Chuang}},
  \emph{\bibinfo{title}{Quantum computation and quantum information}}
  (\bibinfo{publisher}{Cambridge University Press},
  \bibinfo{address}{Cambridge}, \bibinfo{year}{2000}).

\bibitem{garon}
\bibinfo{author}{\bibfnamefont{A.}~\bibnamefont{Garon}},
  \bibinfo{author}{\bibfnamefont{R.}~\bibnamefont{Zeier}}, \bibnamefont{and}
  \bibinfo{author}{\bibfnamefont{S.~J.} \bibnamefont{Glaser}},
  \emph{\bibinfo{title}{Visualizing Operators of Coupled Spins Systems}},
  \bibinfo{journal}{Phys. Rev. A} \textbf{\bibinfo{volume}{91}},
  \bibinfo{pages}{042122} (\bibinfo{year}{2015}).

\bibitem{sjoqvist}
\bibinfo{author}{\bibfnamefont{E.}~\bibnamefont{Sj\"oqvist}},
\emph{\bibinfo{title}{A new phase in quantum computation}},
  \bibinfo{journal}{Physics} \textbf{\bibinfo{volume}{1}}, \bibinfo{pages}{35}
  (\bibinfo{year}{2008}).

\bibitem{geometricgateNA1}
\bibinfo{author}{\bibfnamefont{W.}~\bibnamefont{Xiang-Bin}} \bibnamefont{and}
  \bibinfo{author}{\bibfnamefont{M.}~\bibnamefont{Keiji}},
  \emph{\bibinfo{title}{Nonadiabatic conditional geometric phase shift with NMR}},
  \bibinfo{journal}{Phys. Rev. Lett.} \textbf{\bibinfo{volume}{87}},
  \bibinfo{pages}{097901} (\bibinfo{year}{2001}).

\bibitem{geometricgateNA2}
\bibinfo{author}{\bibfnamefont{S.-L.} \bibnamefont{Zhu}} \bibnamefont{and}
  \bibinfo{author}{\bibfnamefont{Z.~D.} \bibnamefont{Wang}},
  \emph{\bibinfo{title}{Implementation of univeral quantum gates based on nonadiabatic geometric phases}},
  \bibinfo{journal}{Phys. Rev. Lett.} \textbf{\bibinfo{volume}{89}},
  \bibinfo{pages}{097902} (\bibinfo{year}{2002}).

\bibitem{BR1}
\bibinfo{author}{\bibfnamefont{A.} \bibnamefont{Tannus}} \bibnamefont{and}
  \bibinfo{author}{\bibfnamefont{M.} \bibnamefont{Garwood}},
  \emph{\bibinfo{title}{Adiabatic pulses}},
  \bibinfo{journal}{NMR Biomed.} \textbf{\bibinfo{volume}{10}},
  \bibinfo{pages}{423} (\bibinfo{year}{1997}).

\bibitem{BR2}
\bibinfo{author}{\bibfnamefont{M.} \bibnamefont{Garwood}} \bibnamefont{and}
  \bibinfo{author}{\bibfnamefont{Y.} \bibnamefont{Ke}},
  \emph{\bibinfo{title}{Symmetric pulses to induce arbitrary flip angles with compensation for RF inhomogeneity and resonance offsets}},
  \bibinfo{journal}{J. Magn. Reson.} \textbf{\bibinfo{volume}{94}},
  \bibinfo{pages}{511} (\bibinfo{year}{1991}).

\bibitem{khaneja}
\bibinfo{author}{\bibfnamefont{N.} \bibnamefont{Khaneja}},
\bibinfo{author}{\bibfnamefont{B.} \bibnamefont{Heitmann}},
\bibinfo{author}{\bibfnamefont{A.} \bibnamefont{Sp\"orl}},
\bibinfo{author}{\bibfnamefont{H.} \bibnamefont{Yuan}},
\bibinfo{author}{\bibfnamefont{T.} \bibnamefont{Schulte-Herbr\"uggen}},
\bibnamefont{and} \bibinfo{author}{\bibfnamefont{S.~J.} \bibnamefont{Glaser}},
  \emph{\bibinfo{title}{Shortest paths for efficient control of indirectly coupled qubits}},
  \bibinfo{journal}{Phys. Rev. A} \textbf{\bibinfo{volume}{75}},
  \bibinfo{pages}{012322} (\bibinfo{year}{2007}).



\bibitem{VZGS14}
\bibinfo{author}{\bibfnamefont{L.}~\bibnamefont{Van~Damme}},
  \bibinfo{author}{\bibfnamefont{R.}~\bibnamefont{Zeier}},
  \bibinfo{author}{\bibfnamefont{S.~J.} \bibnamefont{Glaser}},
  \bibnamefont{and} \bibinfo{author}{\bibfnamefont{D.}~\bibnamefont{Sugny}},
   \emph{\bibinfo{title}{Application of the Pontryagin maximum principle to the time-optimal
	control in a chain of three spins with unequal couplings}},
  \bibinfo{journal}{Phys. Rev. A} \textbf{\bibinfo{volume}{90}},
  \bibinfo{pages}{013409} (\bibinfo{year}{2014}).

\bibitem{perelomov}
\bibinfo{author}{\bibfnamefont{A.~M.} \bibnamefont{Perelomov}},
 \emph{\bibinfo{title}{Motion of four-dimensional rigid body around a fixed point: An elementary approach}},
  \bibinfo{journal}{J. Phys. A} \textbf{\bibinfo{volume}{38}},
  \bibinfo{pages}{801} (\bibinfo{year}{2005}).

\end{thebibliography}
\end{document}